\documentclass[12pt]{iopart}
\usepackage{graphicx,epsfig}
\usepackage{setstack,iopams,bm}
\begin{document}

\title[Full
counting statistics for noninteracting fermions]{Full
counting statistics for noninteracting fermions:
Exact finite temperature results and generalized long time approximation}
\author{K. Sch\"onhammer}

\address{Physik-Department, Technische Universit\"at M\"unchen, D-85748,
  Garching, Germany\\
Institut f\"ur Theoretische Physik, Universit\"at
  G\"ottingen, Friedrich-Hund-Platz 1, D-37077 G\"ottingen}
\ead{schoenh@theorie.physik.uni-goettingen.de}

\begin{abstract}
Exact numerical results for  the full counting statistics (FCS) of a 
one-dimensional tight-binding model of noninteracting electrons
are presented at finite temperatures using an identity recently
published by Abanov and Ivanov.  A similar idea is used
to derive an explicit expression for the cumulant generating function for a 
system consisting of two quasi-one-dimensional leads connected by a
quantum dot in the long time limit, generalizing the Levitov-Lesovik
formula for two single channel leads to systems with an arbitrary
number of transverse channels.

\end{abstract}

\submitto{\JPCM}
\maketitle

\section{Introduction}

The theory of noise in quantum transport in mesoscopic
systems is a very active field of research \cite{BB,Naza}.
In addition to the first few moments of the transmitted charge 
the full probability distribution can be adressed, called
{\it full counting statistics} (FCS).
The systems usually studied consist of a finite ``dot''-region
connected to $N$ leads which initially are separated from the dot
region and have different chemical potentials \cite{LL1,LL2,Nazarov,Bruder}.
After connecting the subsystems the time evolution of the particle
 transfer between the leads is studied.

In this paper we focus on systems with two quasi-one-dimensional
noninteracting leads. The ``left'' lead consists of $M_L$ tranverse
channels and the initial state is described by a grand canonical
ensemble with chemical potential $\mu_L$ and temperature
$T_L=\beta_L/k_B$. This lead is connected via a finite dot region
to the ``right'' lead with  $M_R=M-M_L$ tranverse
channels, chemical potential $\mu_R \le \mu_L $ and temperature
$T_R=\beta_R/k_B$. 

For noninteracting electrons the calculation of the characteristic 
function of the probability function of the transferred charge can be
exactly reduced to the evaluation a time dependent determinant over the full
one-particle Hilbert space \cite{Klich}. For the lattice systems 
studied in this paper the dimension of this space is finite
before taking the thermodynamic limit.  Exact results
can be obtained for very large but finite systems by numerically
calculating this determinant. 
For times $t$ smaller than the time it takes the ``charge fronts''
which move into the subsystems after connecting them to return to the
connection point after the reflection at the boundaries
 the results are almost independent of the size of the
system \cite{KS}.

 The long time limit  was investigated by Levitov
and Lesovik \cite{LL1} for infinite leads.
 They presented the leading time order result for 
the logarithm of the characteristic function (linear in $t$) as an 
energy integral over the logarithm of a determinant of an
$M \times M$ matrix which
involves the scattering matrix
$s(\epsilon)$ for a single particle. For two single
channel leads, i.e. a strictly one-dimensional system,
 the $2\times 2$ determinant was explicitely evaluated \cite{LL1} .
The result involving the transmission probability $T(\epsilon)$
and the Fermi functions of the two leads is usually called
``Levitov-Lesovik formula''.
 In order to contrast it from the general
leading long time approximation involving the determinant of a
$M\times M$ matrix it is called ``two single channel leads Levitov-Lesovik 
formula'' in the following. For perfect transmission this
formula leads at zero temperature 
to a delta function for the probability distribution of the
transferred charge
incorrectly signalling ``zero shot noise''. 
It was shown previously that in the case of perfect transmission
at zero temperature
the logarithm of the characteristic function 
 increases {\it logarithmically} with time leading to
a probability distribution of finite width also at zero temperature
\cite{MA,KS}.  

 In addition to  exact numerical calculations for $M=2$
an explicit result for the temperature dependent Levitov-Lesovik $M \times M$
determinant is presented in terms of the eigenvalues of a temperature
independent matrix. The result is a sum of terms in the form 
appearing in the two single
channel lead Levitov-Lesovik formula
and might have guessed from the well known results for the first two
cumulants for this general case \cite{BB}. 

For the strictly one-dimensional case $M_L=M_R=1$ exact numerical results were
 obtained for
large but finite lattices
at zero temperature \cite{KS}.
 The first step to obtain 
 the probability distribution of the number of electrons
transmitted to the right lead
was to calculate the time dependence of the one-particle 
projection operator $P_R(t)$ onto the right lead. At zero temperature
only    $\bar P_R(t)= \bar n_0 P_R(t)\bar n_0 $ enters,
where $\bar n_0 $ is the projection operator onto the
initially occupied one-particle states.
 The eigenvalues $p_m(t)$ of $\bar P_R(t) $ determine the probability
distribution $w_R$ of the number of particle transferred to the right lead. 
 The time dependent entanglement entropy after connecting
the subsystems can also be simply expressed in terms of these eigenvalues
\cite{KL}. 

There are approximately $N_t=t (\mu_L-\mu_R)/(2\pi)$
 eigenvalues $p_m(t)\approx T(\mu_R+2\pi(m-1/2)/t) $ different from
zero and one in the long time limit,
where $T(\epsilon)$ is the transmission probability\cite{KS}.
The transition region of the finite eigenvalues to the zero
eigenvalues is not captured by this expression. In order to 
obtain analytical approximations for the exact numerical eigenvalues
in this regime
the logarithmic correction in the large time limit has to be known. 

 For finite temperatures the probability distribution is
determined by $P_R(t)$ in the full one-particle Hilbert space. A
clever rewriting of the determinantal expression
for the characteristic function \cite{AI} allows a simple generalizion
of the zero temperature 
numerical procedure to obtain finite temperature results.
In section 2 exact results are presented for a strictly one-dimensional tight
binding model, i.e. $M_L=M_R=1$.
In the long time limit an accurate analytical approximation for the eigenvalues
of the temperature dependent operator $X(t)$
introduced by Adamov and Ibanov (AI) \cite{AI} is presented which replaces
$\bar P_R(t)$ at finite temperatures.

The analytical expression 
for the eigenvalues of $X(t)$ 
is extendend to arbitrary values of $M_L$
and $M_R$ in section 4 by using 
 a similar rewriting as used by Adamov and Ivanov \cite{AI} for the 
leading order in $t$ result
for the logarithm of  
the characteristic function $g_R$ corresponding to $w_R$. 
This derivation also provides a simple derivation of the 
explicit generalized Levitov-Lesovik formula already mentioned.
For the
special case
$M_R=2$ a comparison is made with analytical results derived
earlier \cite{IS}.

 In section 5
the generalized long time approximation is elucidated for a model
with leads which are stripes of equal width and apart from a single
site impurity
a perfect transition region.  

\section{Counting statistics for noninteracting electrons}

\subsection{General formulation}

\noindent In the following we consider a system which consists
of a finite ``dot''-region described by the Hamiltonian $H_0^{\rm dot}$
connected to the left and right lead with Hamiltonians
$H_{0,a}$ with $a=L,R$. The leads are initially separated from the dot
region . The number of
electrons in the initial state are $N_0^{\rm dot}$ and $N_{0,a}$.
We assume the intial state $|\Phi(0)\rangle$ to be an eigenstate
of $H_0^{\rm dot}$ and the $H_{0,a}$
\begin{equation}
\label{Phi0}
|\Phi(0)\rangle=|E_i^{N_0^{\rm dot}}\rangle \otimes |E_n^{N_{0,L}}\rangle 
\otimes |E_p^{N_{0,R}}\rangle   .
\end{equation} 
The time evolution for times greater than zero is described by
the Hamiltonian 
\begin{equation}
\label{Hamiltonian}
H=  H_0^{\rm dot}+ H_{0,L} + H_{0,R}+ \sum_{a}V_{a}\equiv H_0 +V~.
 \end{equation} 
The term $V$ which couples the leads with the dot region will be specified 
later.
The probability distribution that $Q$ electrons are transferred to
the right system after time $t$ is given by
\begin{eqnarray}
\label{w1}
w_R(t,Q)&=& \langle \Phi(t)|\delta[Q-(
{\cal N}_R-N_{0,R})]|\Phi(t)\rangle  \\
&=&
\frac{1}{2\pi}\int d\lambda e^{-i\lambda Q} g_R(t,\lambda)~. \nonumber
\end{eqnarray}
Here ${\cal N}_R$ ist the particle number operator of the right lead 
and $g_R(t,\lambda) $ is the characteristic function. With
the particle number operators ${\cal N}_a(t) $ in the Heisenberg
picture it is given by
\begin{equation}
g_R(t,\lambda)=
 \langle \Phi(0)|   e^{i\lambda {\cal N}_R(t) }
  e^{-i\lambda {\cal N}_R  }| \Phi(0) \rangle~.
\end{equation} 
The assumption that the initial state is an eigenstate of the
particle number operators was used.
 For initially grand canonical subensembles with different
temperatures and chemical potentials
\begin{eqnarray}
\rho_0^{(a)}= \frac{e^{-\beta_a (H_{0,a}-\mu_a{\cal N}_a)}}
{ {\rm Tr}_a e^{-\beta_a(H_{0,a}-\mu_a{\cal N}_a)} }~,
 \end{eqnarray}
and $\rho_0^{\rm dot} $  of the same type, which corresponds to
a total statistical operator
 $\rho_0$ of the generalized canonical form $\rho_0=e^{-\bar H_0}/\bar Z_0$
the averaging yields for the characteristic function 
\begin{equation}
\label{char}
g_R(t,\lambda)=
\left \langle e^{i\lambda {\cal N}_R(t) }  e^{-i\lambda {\cal N}_{R}}
 \right \rangle~,
\end{equation} 
where $\langle...\rangle$ denotes the averaging with the statistical
operator $\rho_0$. This result is also valid for interacting fermions.\\

\subsection {Noninteracting fermions}

For noninteracting fermions the characteristic
function can be expressed as a determinant in the one particle Hilbert
space using Klich's trace formula\cite{Klich,KS}
\begin{eqnarray}
\label{Kli}
g_R(t,\lambda)&=& {\rm det}\left [ 1 +\left (e^{i\lambda P_R(t)}
e^{-i\lambda P_R}  - 1\right )\bar n_0
  \right  ] \\
&\equiv & {\rm det}\left [1+a(t)\bar n_0\right ]=
{\rm det}\left [1+ \bar n_0 a(t)\right ]~,\nonumber
\end{eqnarray}
where $\bar n_0=(e^{\bar
  h_0}+1)^{-1}$ is the Fermi operator. It is determined by
the Fermi functions describing the initial state.
The equality in the second line holds because the inverse of $\bar n_0$ 
exists, in
contrast to the zero temperature case where $\bar n_0$ is a projection
 operator. 
 
Using  $e^{i\lambda P_R(t)}= 1 +(e^{i\lambda}-1) P_R(t) $ 
and the definition $d(\lambda)=e^{i\lambda}-1$ the
operator $1+\bar n_0 a$ can be written in the form proposed by AI \cite{AI} 
\begin{eqnarray}
\label{AItrick}
1+\bar n_0 a&=&\left[ e^{i\lambda P_R}+\bar n_0 \left(    
e^{i\lambda P_R(t)} - e^{i\lambda P_R}  \right )    \right
]e^{-i\lambda P_R} \nonumber \\
&=& \left\{1+d(\lambda)\left[(1-\bar n_0)P_R+\bar n_0 P_R(t)\right ]
 \right\} e^{-i\lambda P_R}  \nonumber \\
&\equiv& \left [ 1+d(\lambda)X(t)     \right]e^{-i\lambda P_R}~.
\end{eqnarray}
As  $ \bar n_0^{-1/2}$
exists and $ \bar n_0$ and $P_R$ commute it is 
 more convenient to work with 
$\tilde X(t)=\bar n_0^{-1/2} X(t)\bar n_0^{1/2} $ i.e.
\begin{equation}
\label{Xtilde}
\tilde X(t)=(1-\bar n_0)P_R+\bar n_0^{1/2}P_R(t)\bar n_0^{1/2}~.
\end{equation}
 This yields for the characteristic 
function for arbitrarily large but finite systems
\begin{eqnarray}
\label{AbIv}
g_R(t,\lambda)&=&e^{-i\lambda N_R}{\rm det}\left [1+d(\lambda)\tilde
  X(t)\right] \nonumber \\
&=&
 e^{-i\lambda N_R}\prod_{m=1}^{N_H}
 \left [1+(e^{i\lambda}-1)\tilde X_m(t)\right ] \nonumber \\
&=& \sum_{n=0}^{N_H}c_n(t)e^{i(n-N_R)\lambda}~,
\end{eqnarray}
where $N_H$ is the dimension of the total one-particle Hilbert space
and $N_R $ of the one
of the right lead.
They are both finite for finite lattice systems.
The coefficients $c_m(t)$ can be obtained recursively from the
eigenvalues $\tilde X_n(t)$ of the AI one-particle operator $\tilde X(t)$
defined in Eq. (\ref{Xtilde}) as described shortly in the appendix.

 Apart from the replacements
$N_{0,R} \to N_R, N_{\rm tot}\to N_H$ and $p_m(t)\to \tilde
X_m(t)$
 this finite temperature result has the
same form as the $T=0$ approach which was used as the starting point for
the exact numerical calculation of the FCS \cite{KS}.
 The probability distribution at finite temperatures
is given by 
\begin{equation}
w_R(t,Q)= \sum_{n=1}^{N_H}c_n(t)\delta\left
  (Q-(n-N_R)\right )~.
\end{equation}
To obtain exact results for the FCS one first has to calculate
$\tilde X(t)$ using the result for
 $P_R(t)$ and then obtain its eigenvalues $\tilde X_m(t)$.
This is done for the simplest case  $M_L=M_R=1$ in the following section.

\section{Exact results  for  $M_L=M_R=1$ }

\subsection{The model}

In this section we present exact numerical results for the
probability distribution $w(t,Q)$ for a one dimensional tight
binding model with a one site (noninteracting) dot.
 The unperturbed one-particle Hamiltonians of the subsystems
 are given by
\begin{eqnarray}
\label{hamiltonian}
h_{0,a}&=&-t_{\|} \sum_{m=1}^{N_a-1} (|am\rangle\langle a
(m+1)|+H.c.),\nonumber \\
h_0^{\rm dot}&=& V_0|0\rangle \langle 0|
\end{eqnarray}
The number of sites in the leads are given by $N_a$. In
$|am\rangle $
the label $a$ takes the value $1$ for $a=R$ and $-1$ for $a=L$.
The hopping matrix elements in the leads $t_{\|}$ are taken as unity
in the numerical calculations which leads to  total bandwidth of $4$.
The eigenstates of the unconnected leads are standing waves.
 The coupling between the subsystems is described by the
hopping term
\begin{equation}
\label{connection}
v=-t_L |-1\rangle\langle 0|
-t_R |0\rangle\langle 1|+H.c.
\end{equation}

\subsection{ Numerical results}

The first step to calculate  the eigenvalues $\tilde X_m(t)$ is to
obtain $P_R(t)$ using the time dependence of the one-particle states
 $|\epsilon_\alpha^{(0)}\rangle $  
\begin{equation} 
\langle \epsilon_\alpha^{(0)}|P_R(t) |\epsilon_\beta^{(0)}\rangle
=\sum_{m=1}^{N_R} \langle \epsilon_\alpha^{(0)}(t)|m\rangle \langle
 m|\epsilon_\beta^{(0)}(t)\rangle~.
\end{equation}
The time dependence of the states is calculated using the spectral
decomposition of the full one-particle Hamiltonian \cite{KS}.
From the  resulting $N_H\times N_H$ matrix one obtains $\tilde X(t)$
as prescribed in Eq. (\ref{Xtilde}).
In the following we show results for identical temperatures in the
initial subsystems and $\mu_L=\mu_{\rm dot}>\mu_R$.

 We begin with a generic example $t_L=0.8, t_R=0.5$
and $V_0=0.4$. As the coupling of the dot to the leads is asymmetric
the transmission probability at the resonance energy is less then one.
The results shown are for time $t=200$. If twice the number of lead sites $N_a$
is larger than
$v_{\rm max}t=2t$  the results
for the eigenvalues  $\tilde X_n$ which differ from zero and one
 become independent of the 
$N_a$. For the exact numerical results shown we
used $N_L=N_R=500$. As $\tilde X(0)=P_R$ 
holds there are $N_R$
eigenvalues one and $N_L+1$ eigenvalues zero
 at the initial time. Therefore we show the eigenvalues in descending
order as a function of $n-N_R$.
 In Fig. 1 the eigenvalues $\tilde X_n$ are
shown for three different temperatures.
  \vspace{0.5cm}    
\begin{figure}[tb]
\begin{center}
\vspace{-0.0cm}
\leavevmode
\epsfxsize8.0cm
\epsffile{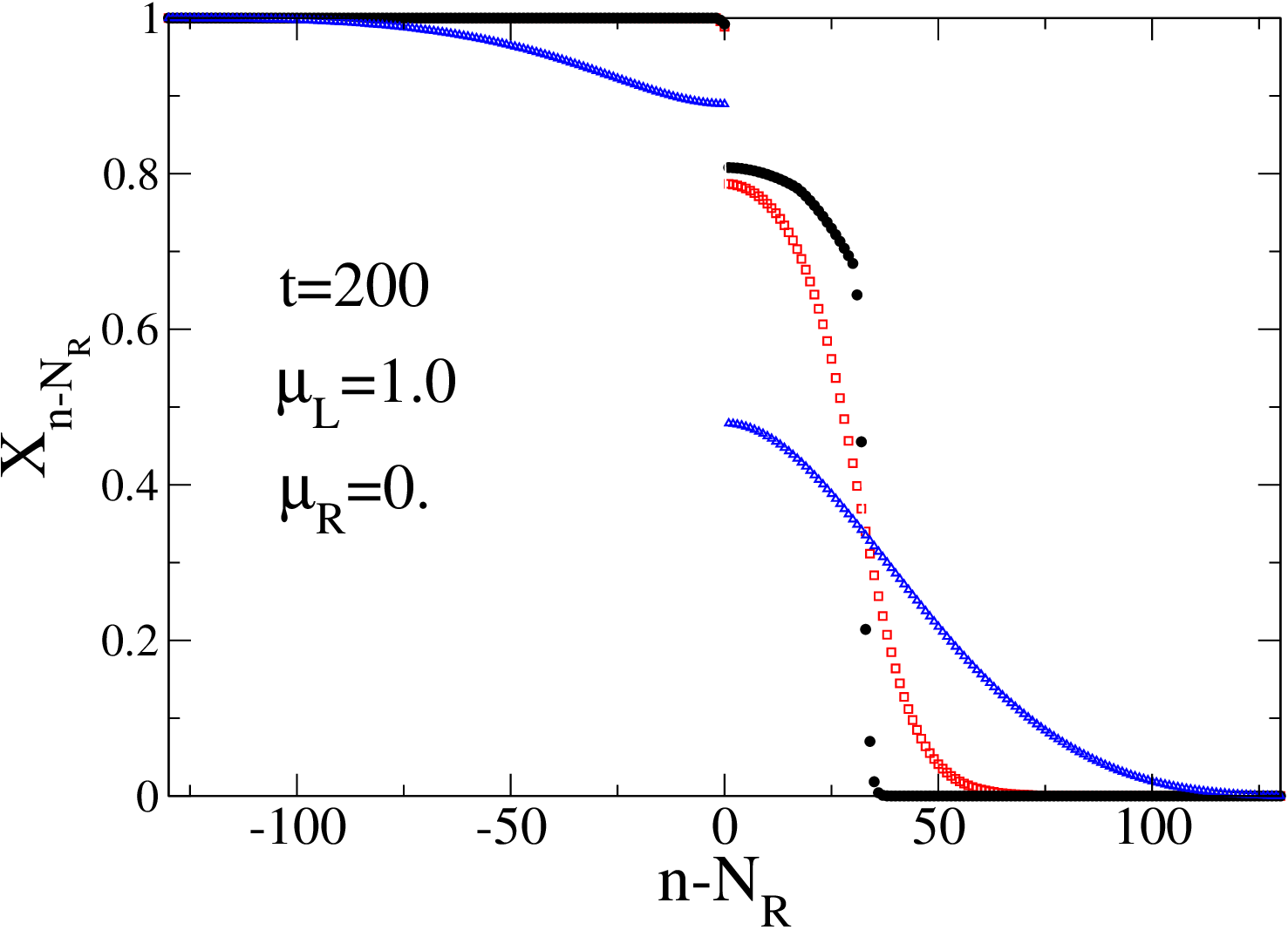}
\caption {Eigenvalues of $\tilde X(t) $ 
which differ from one and zero for $t_L=0.8, ~t_R=0.5$
and $V_0=0.4$ at time $t=200$ for three different temperatures, i.e.
values of $\beta$. The full circles correspond to $\beta=2000$ which
hardly differs from the zero temperature result. The open 
 squares are the results for $\beta=10$ and the open triangles for
 $\beta=2$. }
\label{EW1}
\end{center}
\end{figure}

As mentioned in the introduction the ``zero temperature'' result
$\beta=2000$ can be well described analytically except for the narrow
transition region to the zero eigenvalues which is related to the 
logarithmic corrections in the long time limit \cite{KS,MA}.
For $\beta=10$ the main effect is to smooth out this transition
region. A new effect sets in at larger temperatures.
For $\beta=2$ part of the 
eigenvalues with  $n<N_R$  which are one at the initial time 
 get visibly reduced.

In order to understand this behaviour
analytically we start from the Levitov-Lesovik formula \cite{LL1,LL2}
\begin{equation}
\label{Integral}
\ln g_R(t,\lambda)=\frac{t}{2\pi}\int_{-B}^B\ln
(1+F(\epsilon,\lambda))d\epsilon~,
\end{equation}
where $B=2$ for the choice $t_{\|}=1$ and
\begin{equation}
\label{Flambda}
F(\epsilon,\lambda)=T(\epsilon)\left (d(\lambda)f_L(\epsilon)\bar f_R(\epsilon)
+ d^*(\lambda)f_R(\epsilon)\bar f_L(\epsilon)\right )~,
\end{equation}
whith $\bar f_a\equiv 1-f_a$.
We approximate the integral in Eq. (\ref{Integral}) by a finite
Riemann sum over $N$ intervals of size $2B/N$ and use the trapecoidal
rule
\begin{equation}
\label{Riemannsum}
\ln g_R \approx \frac{t}{2\pi} \frac{2B}{N} \sum_{j=1}^N
\ln \left [1+F\left (-B+(j-\frac{1}{2})\frac{2B}{N},\lambda\right )\right ]~,
\end{equation}
which agrees with the integral in the limit $N\to \infty$. As
Eq. (\ref{Integral}) is itself an approximation for the large time
limit we choose $N=N(t)$ with
\begin{equation}
N(t)=\frac{t}{2\pi} 2B~.
\end{equation}
Then the prefactor in the sum equals one and $g_R(t)$ takes a form
which can easily be compared with Eq.(\ref{AbIv})
\begin{eqnarray}
 g_R(t,\lambda)\approx \prod_{j=1}^{N(t)}
\left [1+F (\epsilon_j,\lambda)\right ]~,
\end{eqnarray}
where $\epsilon_j\equiv -B+2\pi(j-1/2)/t $.

In the low temperature regime $k_BT_a \ll \mu_L-\mu_R$ the
approximation for the eigenvalues $\tilde X_j$ can easily be read off
as the factor $f_R\bar f_L$ multiplying $d^*$ 
in $F(\epsilon,\lambda) $ in Eq. (\ref{Flambda})
is exponentially small.
With $e^{-i\lambda}(1+d(\lambda))=1$ the comparison with Eq.(\ref{AbIv})
shows that the eigenvalues different from one are given by
\begin{equation}
\label{approx}
\tilde X_j\approx T(\epsilon_j) f_L(\epsilon_j)\bar f_R(\epsilon_j)
\approx  T(\epsilon_j) (f_L(\epsilon_j)-f_R(\epsilon_j))
\end{equation}
At zero temperature this agrees with the result mentionend in the 
introduction \cite{KS}.

At arbitrary temperatures one has to factor  $1+F(\epsilon,\lambda) $
 with $F$ defined in Eq. (\ref{Flambda}) in the form
\begin{equation}
1+F(\epsilon,\lambda)=e^{-i\lambda}(1+a_+(\epsilon)d(\lambda))
(1+a_-(\epsilon) d(\lambda))
\end{equation}
The comparison with Eq. (\ref{Flambda}) yields
\begin{equation}
\label{EWapp}
a_{\pm}=\frac{1+T(f_L-f_R)}{2}\pm w
\end{equation}
with 
\begin{eqnarray}
\label{diff}
w&=&
\sqrt{ \left ( \frac{1+T(f_L-f_R)}{2}\right  )^2-Tf_L\bar f_R}\\
&=&
\sqrt{ \left ( \frac{1-T(f_L-f_R)}{2}\right  )^2-Tf_R\bar
  f_L}\nonumber \\
&=&
\sqrt{ \left ( \frac{1-T(f_L+f_R)}{2}\right  )^2+T(1-T)f_Lf_R}~.\nonumber
\end{eqnarray}
The second form for $w$ is useful for the
discussion of the low temperature results
and the third form shows that there is an energy gap in the spectrum for
non-perfect transmission.
It also shows that the factorization is simplest for perfect
transmission
\begin{equation}
\label{perfectfac}
1+F(\epsilon,\lambda)=e^{-i\lambda}(1+f_L(\epsilon)d(\lambda))
(1+\bar f_R(\epsilon) d(\lambda))~.
\end{equation}

\vspace{0.5cm}    
\begin{figure}[tb]
\begin{center}
\vspace{-0.0cm}
\leavevmode
\epsfxsize8.0cm
\epsffile{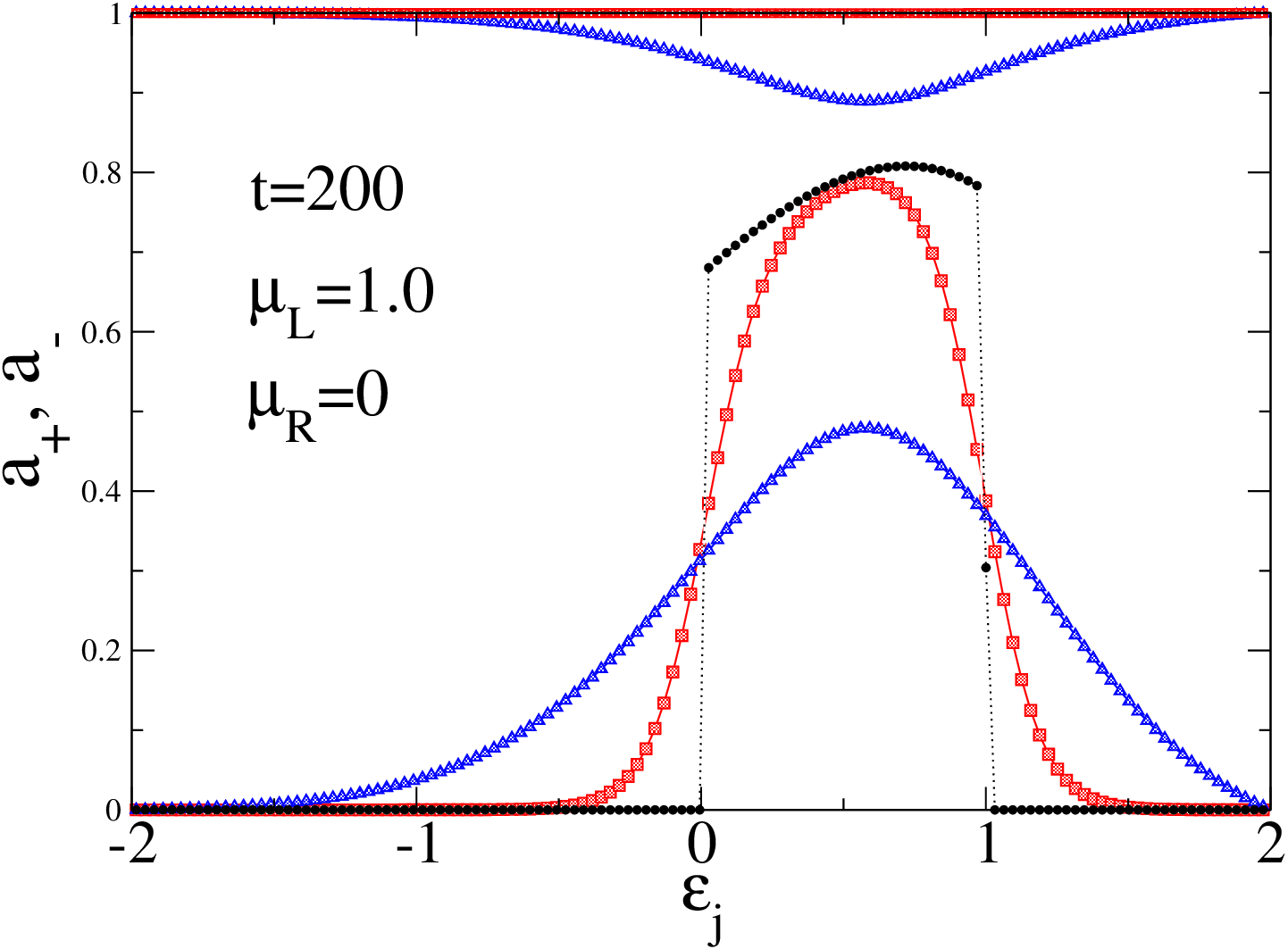}
\caption { Long time approximation
Eq. (\ref{EWapp}) for the eigenvalues $\tilde X(t) $
plotted as a function of the $\epsilon_j$ for the parameter values of
Fig.1 . Filled dots: $\beta=2000$, open squares: $\beta=10$, open
triangles: $\beta=2$. }
\label{EW2}
\end{center}
  \vspace{0.5cm} 
\end{figure}

In Fig. 2 we show the analytical approximation Eq. (\ref{EWapp})
 for the eigenvalues
 for the parameter values used in Fig. 1 as a function
of the $\epsilon_j$. The result for $\beta=2000$
 (filled dots)
is almost identical to the zero temperature results. The eigenvalues
$a_-(\epsilon_i)$
are nonzero in the energy range between $\mu_R$ and $\mu_L$ and show
the energy dependence of the transmission probability. The eigenvalues
 $a_+(\epsilon_i)$ are almost identical to one. Also for $\beta=10$
(open squares) the
 $a_+(\epsilon_i)$ equal one within the drawing accuracy while the
$a_-(\epsilon_i)$ are very well approximated by Eq. (\ref{approx}) and
show the thermal broadening of the zero temperature result. For
$\beta=2$ (open triangles) Eq. (\ref{approx})  no longer presents
 a good approximation  for the  $a_-(\epsilon_i)$ and the
are $a_+(\epsilon_i)$ are clearly smaller than one in a large energy
range.

\begin{figure}[tb]
\begin{center}
\vspace{0.5cm}
\leavevmode
\epsfxsize8.0cm
\epsffile{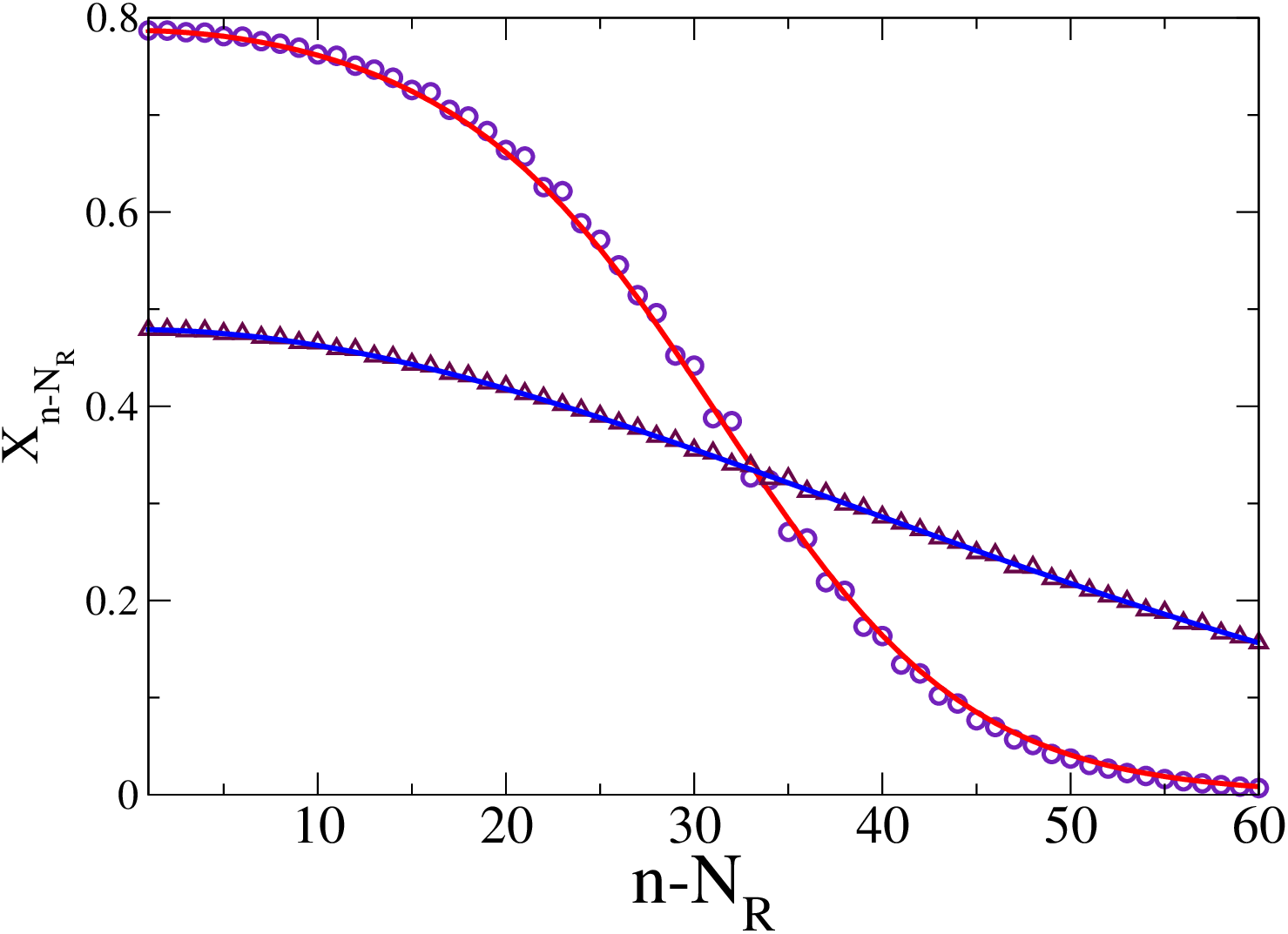}
\caption{Comparison of the exact eigenvalues of Fig.1 (shown as the
  full lines) with the approximations $a_-$ presented in descending
  order. The results for $\beta=10$ and $2$ 
are shown in a restricted range.}
\label{EW3}
\end{center}
\end{figure}
\vspace{1.0cm}

In order to compare the approximate eigenvalues $a_\pm$ with the exact
numerical eigenvalues $\tilde X_n$ the  $a_\pm$ have to be brought
into descending order. This is shown in Fig. 3.
For $\beta=2$ the $a_-$ (open triangles) agree very well with the
exact results apart from the fact that there are pairs of almost equal
eigenvalues. As shown later this has a rather small effect for the 
calculation of the probability distribution $w_R$. For $\beta=10$ the
deviations are a bit larger. As shown earlier
the transition region to the zero eigenvalues for $\beta=2000$ 
 is not captured by the
approximation $T(\epsilon_i)$\cite{KS}.

\vspace{0.5cm}
\begin{figure}[tb]
\begin{center}
\vspace{0.5cm}
\leavevmode
\epsfxsize8.0cm
\epsffile{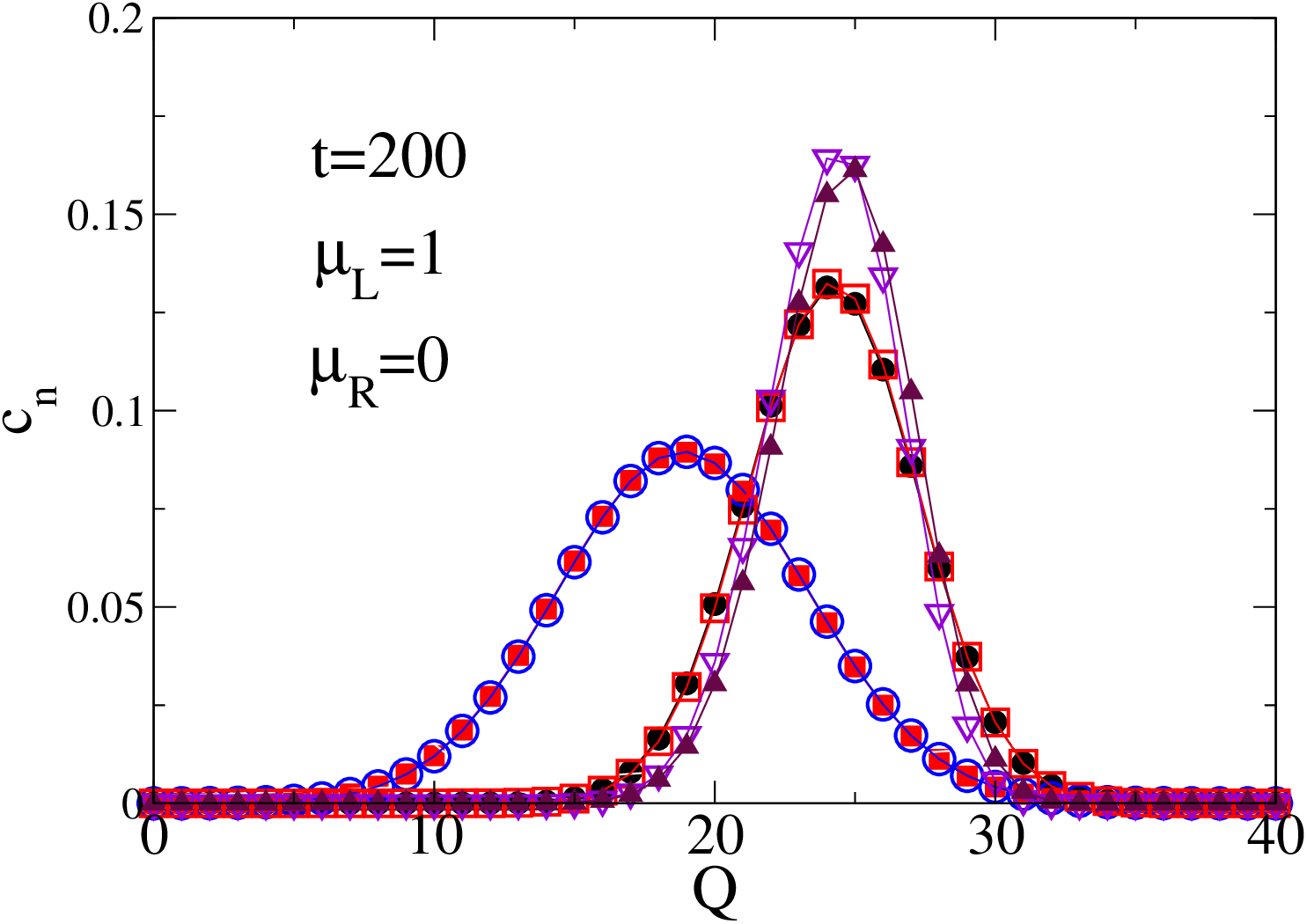}
\caption{ Weights of the probability distribution $w_R(t,Q)$ for the
  parameters
of Fig. 1. The exact results are presented by the full symbols, the
approximation
using Eq. (\ref{EWapp}) by the open ones: $\beta=2$: filled squares,
open circles, $\beta=10$: filled circles, open squres, 
$\beta=2000$: filled triangles, open triangles.}
\end{center}
\end{figure}

In order to obtain the probability distribution $w_R(t,Q)$ form the
eigenvalues $\tilde X_n$ the product in Eq. (\ref{AbIv}) has to be
evaluated recursively. The details are presented in the appendix.  
The results for the parameters of Fig. 1 are shown in Fig. 4.
For $\beta=2$ the exact result (full squares) agrees very well with
the one using the approximation of Eq. (\ref{EWapp}) (open squares).
 For $\beta=10$
the agreement of the exact result (full circles) with the
approximation (open squares) is still rather good. For $\beta=2000$
 the filled and open symbols no longer overlap.

The discrepancy between the exact results and the approximation 
  Eq. (\ref{EWapp}) is most prominent for perfect transmission in
the zero temperature limit. While the Levitov-Lesovik formula predicts
a single delta function of weight one (``zero shot noise'') the
exact result clearly has a finite width\cite{KS} as shown in
Fig. 5 where the $t_L=1$, $t_R=1$ and $V_0=0$ was used leading to 
perfect transmission. 

\vspace{1.0cm}
\begin{figure}[tb]
\begin{center}
\vspace{-0.0cm}
\leavevmode
\epsfxsize8.0cm
\epsffile{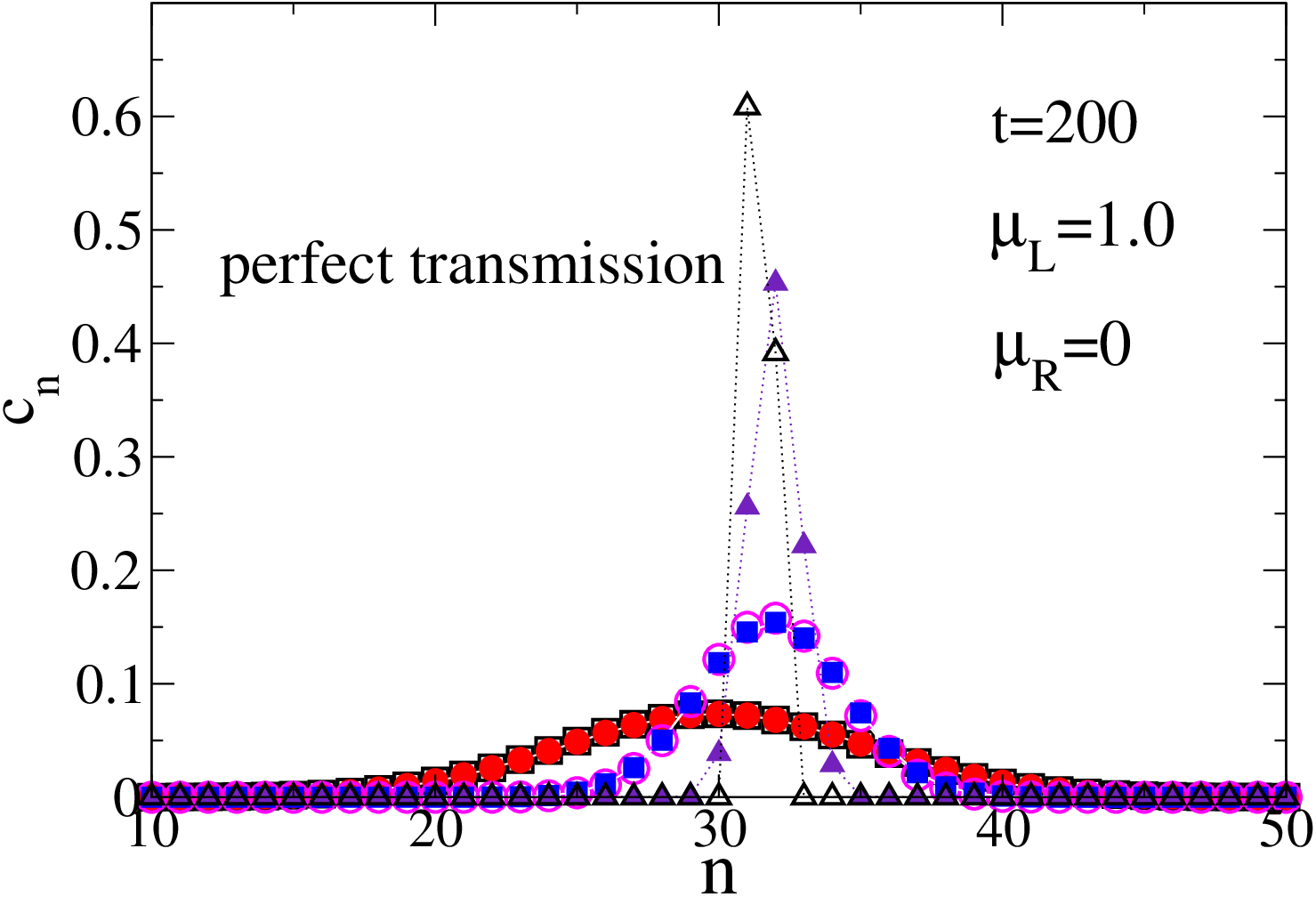}
\caption{ Weights of the probability distribution $w_R(t,Q)$ for the
 case of perfect transmission. The exact results are presented by the full symbols, the
approximation
using Eq. (\ref{EWapp}) by the open ones: $\beta=2$. filled circles,
open squares, $\beta=10$: filled squares, open circles, $\beta=2000$:
filled triangles, open triangles. }
\end{center}
\end{figure}
\vspace{1.0cm}

\subsection {Perfect transmission}

As the discrepancy for very low temperatures
 concerns the width of the approximate
distribution shown in Fig. 5  it is useful
to dicuss the behaviour of the second order cumulant
\begin{eqnarray}
\label{kappa2}
\kappa_2(t)&=&\sum_{n=1}^{N_H}\tilde X_n(t)(1-\tilde X_n(t)) \\
 &\approx&\sum_{j=1}^{N(t)}\left [ a_+(\epsilon_j)(1- a_+(\epsilon_j ))
+ a_-(\epsilon_j)(1- a_-(\epsilon_j )     \right ] \nonumber \\
&=&\sum_{j=1}^{N(t)}\left [T^2(f_L\bar f_L+f_R\bar f_R) 
+T(1-T)(f_L\bar f_R+f_R\bar f_L) \right ]~. \nonumber
\end{eqnarray}
The (leading time order) shot noise contribution \cite{Leso}
 proportional to $T(1-T)$ vanishes for 
perfect transmission and the remaining term is well known in the
limit where the sum is replaced again by the integral \cite{BB,Naza,LL2}.

\vspace{0.5cm}
\begin{figure}[tb]
\begin{center}
\vspace{-0.0cm}
\leavevmode
\epsfxsize8.0cm
\epsffile{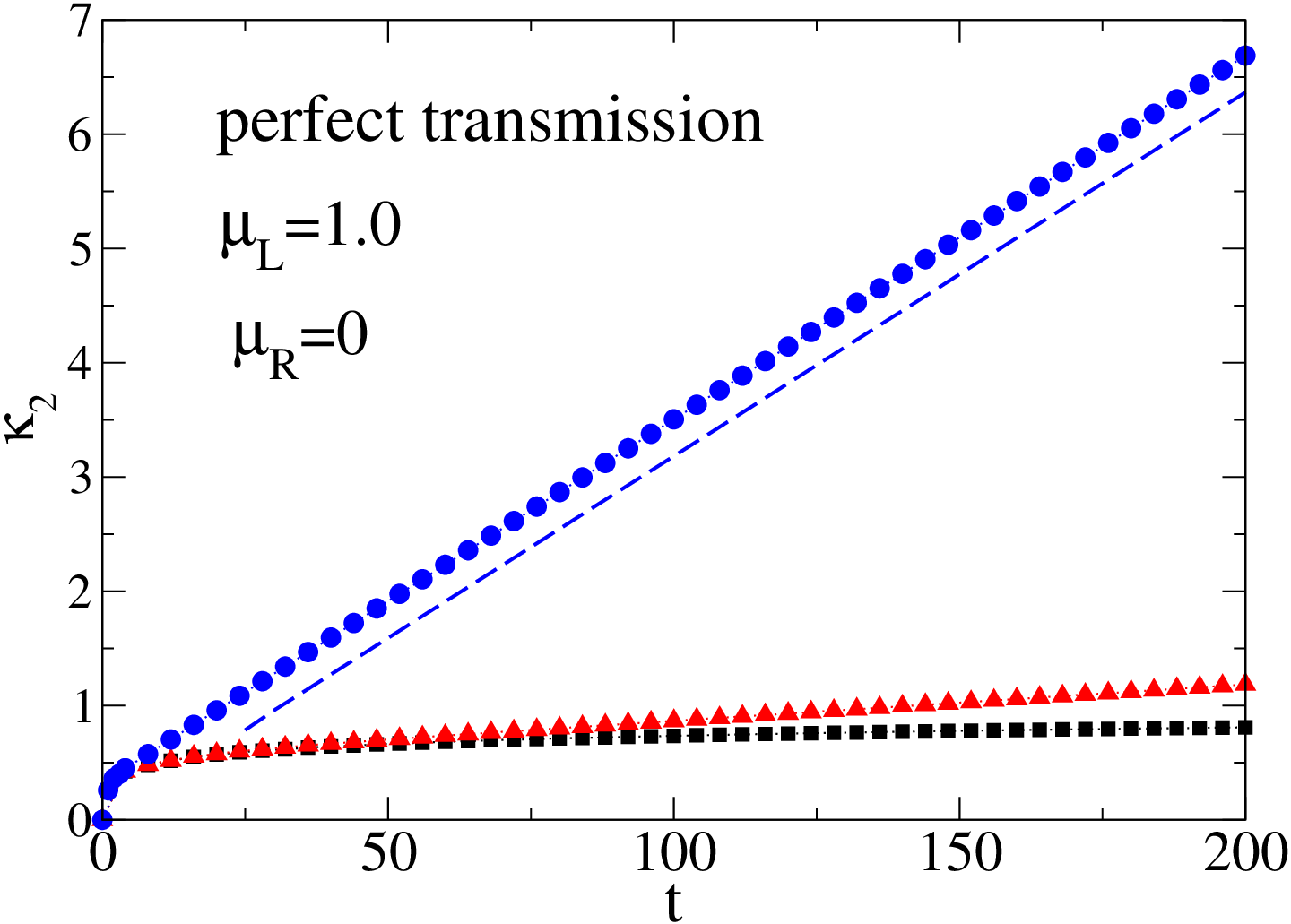}
\caption{Exact results (filled symbols) for second order cumulant
  $\kappa_2$
 for perfect transmission as
  a function of time for three different temperatures: $\beta=2000$:
  filled squares, $\beta=100$: filled triangles and $\beta=10$: filled
  circles.
The dotted line shows the linear part $t/(\pi\beta)$ for $\beta=10$.  }
\label{wRV}
\end{center}
\end{figure}
\vspace{1.0cm}

Because of the factorization in Eq. (\ref{perfectfac}) of $1+F$
for perfect transmission the derivative of $\ln g_R$ with respect to
$i\lambda$ takes the simple form 
\begin{eqnarray}
\label{pt}
\frac{2\pi}{t}     \frac{\partial \ln g_R(t,\lambda)}{\partial(i\lambda)}&=&
\mu_L-\mu_R+\left(    \frac{1}{\beta_R}+   \frac{1}{\beta_L}
\right) i\lambda  \\
&+& \sum_a \frac{1}{\beta_a}\ln\left[     \frac{1+e^{-\beta_a(B-a\mu_a)-i\lambda }}
  {1+e^{-\beta_a(B+a\mu_a)+i\lambda }}     \right ]   \nonumber  ~, 
\end{eqnarray}
where the factor $a$ in $a\mu_a$ again takes the value $1$ for $a=R$
and $-1$ for $a=L$. 
In the wide band limit $\beta_a(B-|\mu_a|)\gg 1$ the logarithmic terms
can be neglected and $w_R$ is a Gaussian in the Levitov-Lesovik
approximation with a temperature independent mean value \cite{LL2} .
The fact that the average charge transfer for $\beta=2$ in Fig. 5 is slightly
less than for $\beta=10$ and $\beta=2000$ is related to the
correction term in Eq.(\ref{pt}) which adds a constant contribution
for $\lambda=0$.

In the wide band limit $\ln g_R(t,\lambda) $ can be calculated
analytically also for an energy independent transmission probability
which differs from one \cite{Inhester}.

Exact numerical results for $\kappa_2$ are shown in Fig. 6 as a function of time
for system size $N_L=N_R=500$ up to times were the result is
independent of this system size. For $\beta=10$ (filled circles) there
is an almost linear increase of $\kappa_2$ rather quickly. The dotted
 line shows the
linear increase which follows from the finite temperature
Levitov-Lesovik formula (see Eq. (\ref{pt})). For $\beta=100$ (filled
triangles)
there is crossover from a logarithmic increase to a linear time
dependence around  $t \approx 50$. For $\beta=2000$ the times shown are
too small to see the corresponding crossover and the shown 
increase is logarithmic
in time.  This logarithmic behaviour \cite{KS,MA,Bla}
is {\it not} captured
by the approximate eigenvalues $a_\pm$ in Eq. (\ref{EWapp}).

\section{Long time behaviour for general values 
of  $M_L$ and $M_R$}

To study the long time limit one first takes the thermodynamic limit
in which the energies of the unperturbed eigenstates $|\epsilon,a,i\rangle$
with $i=1,...,M_a$ of the left and right lead form $M_L$ and $M_R$
continua extending from $\epsilon_{\rm min}^{a,i}$ to  $\epsilon_{\rm
  max}^{a,i}$ which
usually partially overlap. For a given energy $\epsilon$ the 
``open scattering channels''\cite{Taylor} in the leads
are those for which $\epsilon \in [\epsilon_{\rm min}^{a,i},\epsilon_{\rm
  max}^{a,i}   ]$. Their number is denoted as
$M_a(\epsilon)\le M_a$. For a given $\epsilon$ we order the channels
with $i=1,...,M_a(\epsilon)$ as the open channels and  $i=M_a(\epsilon)+1,...,M_a$
the closed channels.
The dimension of the scattering matrix $s(\epsilon) $  for a
single particle \cite{Taylor} is given by
 $M(\epsilon) =M_L(\epsilon)+M_R(\epsilon)\le M$, with
 $M_a(\epsilon)\ge 1$ in order to have scattering at all.

 In the following we
 use a Dirac notation in the $M(\epsilon)$-dimensional space 
with the orthonormal basis $|a,i)$ where the $i$ run from $1$ to
$M_a(\epsilon)$. Then the projection operators $\tilde P_a(\epsilon)$
on the lead channels read
\begin{equation}
\tilde P_a(\epsilon)=\sum_{i=1}^{M_a(\epsilon)}|a,i)(a,i|~.
\end{equation}
The energy dependence of the $\tilde P_a(\epsilon)$ is only via the 
 selection of the open channels.\\

For the general geometry discussed in section 2
 the long time linear in $t$ contibution to $\ln g_R$
takes the form\cite{LL1,IS}  
\begin{equation}
\label{Integralgen}
\ln g_R(t,\lambda)=\frac{t}{2\pi}\int_{\epsilon_{\rm
    min}}^{\epsilon_{\rm max}}\ln
{\rm det}[1+c(\epsilon,\lambda)]d\epsilon~.
\end{equation}
Here $c(\epsilon,\lambda) $ is the $M(\epsilon)\times M(\epsilon)$ matrix 
\begin{equation}
c(\epsilon,\lambda)=\left [ s^\dagger
(\epsilon)e(\lambda,\epsilon )s(\epsilon)e^\dagger(\lambda,\epsilon)
-1\right ]f(\epsilon)~,
\end{equation}
with
\begin{eqnarray}
e(\lambda,\epsilon)&=& \tilde P_L(\epsilon)+e^{i\lambda}\tilde P_R(\epsilon)
=1+d(\lambda)\tilde P_R(\epsilon) \nonumber\\
f(\epsilon)&=& f_L(\epsilon)\tilde P_L(\epsilon)+ f_R(\epsilon) \tilde P_R(\epsilon)~.
\end{eqnarray}
The values of $\epsilon_{\rm min}$ and $\epsilon_{\rm max}$ depend on the details of the
microscopic model.\\
 
Now one can essentially repeat the steps used to derive the AI-form 
Eq. (\ref{AbIv}) to obtain
\begin{eqnarray}
\label{determinant}
{\rm det} [1+c(\epsilon,\lambda)] =e^{-i\lambda M_R(\epsilon)}{\rm det}
\left [1+d(\lambda)\tilde
  X(\epsilon) \right]
\end{eqnarray}
with
\begin{eqnarray}
\label{xtilde}
\tilde X=(1-f)\tilde P_R+f^{1/2}s^\dagger \tilde P_R sf^{1/2}~.
\end{eqnarray}
The determinant in Eq. (\ref{determinant}) can be calculated by first
solving the eigenvalue problem for  $\tilde X(\epsilon)$.
As the $M(\epsilon) \times M(\epsilon)$-matrix
 $\tilde X(\epsilon)$ in Eq.(\ref{xtilde}) is
temperature dependent it looks as if one has to solve a different
eigenvalue problem for each temperature. In the following we show that
this is {\it not} necessary. In fact it is sufficient to solve a {\it
  single} $M_R(\epsilon) \times M_R(\epsilon)$
 eigenvalue problem to obtain a new 
generalized long time approximation for arbitrary temperatures.
 To
show this we write 
\begin{equation} 
\tilde X(\epsilon)=\sum_{j=1}^{M_R(\epsilon)}\left [\bar f_R|R,j)(R,j|+|b_j)(b_j|   \right]
\end{equation} 
with 
\begin{equation} 
|b_j)=f^{1/2}s^\dagger|R,j)~,~~~~(b_j|=(R,j|sf^{1/2}~.
\end{equation} 
Multiplying the eigenvalue problem $\tilde X|X_\alpha)=X_\alpha|X_\alpha)$
from the left with $(b_i|$ yields
\begin{equation} 
\label{first}
\sum_{j=1}^{M_R(\epsilon)}\left[ (b_i|R,j)(R,j|X_\alpha)\bar f_R
+(b_i|b_j)(b_j|X_\alpha)\right]=X_\alpha(b_i|X_\alpha)  
\end{equation} 
and multiplying with 
 $(R,i|$ gives for the overlaps  $(R,i|X_\alpha)$
\begin{equation} 
(R,i|X_\alpha)=\frac{1}{X_\alpha-\bar f_R}\sum_{j=1}^{M_R(\epsilon)}(R,i|b_j)(b_j|X_\alpha)~.
\end{equation}
Inserting this into Eq. (\ref{first}) leads after multiplication with
$ X_\alpha-\bar f_R$ to
\begin{eqnarray} 
\sum_{j=1}^{M_R(\epsilon)}\left[\bar f_R(b_i|\tilde P_R|b_j)+(X_\alpha-\bar f_R)
  (b_i|b_j)\right ] (b_j|X_\alpha) \nonumber \\
= (X_\alpha^2-\bar f_R X_\alpha)(b_i|X_\alpha)~.
\end{eqnarray}
This equation can be rewritten using 
\begin{equation}
(b_i|b_j)=f_R\delta_{ij}+(f_L-f_R)(R,i|s\tilde P_Ls^\dagger|R,j)~.
\end{equation}
Therefore a single hermitian operator in the $M_R(\epsilon)$-dimensional
 subspace spanned by
the $|i\rangle\equiv |R,i)$ determines the original eigenvalue problem
\begin{equation}
\label{xa}
\left[(f_L-f_R)X_\alpha -f_L\bar f_R
\right]\tau|Y_\alpha^{(R)}\rangle
=(X_\alpha^2-X_\alpha)|Y_\alpha^{(R)}\rangle,
\end{equation}
where $ |Y_\alpha^{(R)}\rangle\equiv \tilde P_Rsf^{1/2}|X_\alpha) $ and
\begin{equation}
\label{deftau}
\tau=\tilde P_Rs\tilde P_Ls^\dagger \tilde P_R=
 (\tilde P_Rs\tilde P_L)(\tilde P_Rs\tilde P_L)^\dagger \equiv AA^\dagger~.
\end{equation}
The elements of the matrix $A$ are the
left to right transmission amplitudes \cite{comment}.

The matrix elements of $\tau$ are given by ($i,j\in [1,M_R(\epsilon)]$)
\begin{equation}
\tau_{ij}(\epsilon)=\sum_{l=M_R(\epsilon)+1}^{M(\epsilon)} 
s_{il}(\epsilon)s_{lj}^\dagger(\epsilon)~.
\end{equation}
After solving the eigenvalue problem $\tau|\tau_\mu\rangle
 =\tau_\mu|\tau_\mu\rangle$ the determination of the $X_\alpha$ in
 Eq. (\ref{xa}), after
multiplying with $\langle \tau_\mu|$, is reduced to solving a quadratic equation.
 With $\alpha \to \mu,\pm$ the
solution reads
\begin{equation}
X_{\mu,\pm}=\frac{1+\tau_\mu(f_L-f_R)}{2}\pm
 \sqrt{ \left ( \frac{1+\tau_\mu(f_L-f_R)}{2}\right  )^2-\tau_\mu f_L\bar f_R}~.
\end{equation}
This is like Eqs. (\ref{EWapp}) and (\ref{diff}) with $T(\epsilon_i)$ replaced by
$\tau_\mu(\epsilon_i)$. 

For $M_R(\epsilon)=1$ there exists only one eigenvalue
$\tau_1$, which for $M_L(\epsilon)=1$ is given by $|s_{12}(\epsilon)|^2=T(\epsilon) $.
For $M_R(\epsilon)=1$ and arbitrary values of $M_L(\epsilon)$ the single eigenvalue is
given
\begin{equation}
\tau_1(\epsilon)=\sum_{l=2}^{M_(\epsilon)} |s_{1l}(\epsilon)|^2~.
\end{equation}
This corresponds to the simplest generalization of the Levitov-Lesovik
formula\cite{KS}.

 For $M_R(\epsilon)=2$ the two eigenvalues of $\tau(\epsilon)$ are given by
\begin{equation}
\tau_{1,2}=\frac{{\rm tr}\tau}{2}\pm \sqrt{ \left(\frac{{\rm
      tr}\tau}{2}\right)^2+{\rm det}\tau }~.
\end{equation}
Zero temperature results for this case were presented earlier
\cite{IS} using a different derivation. For the special case $M_L(\epsilon)=1$
the determinant of $\tau$ vanishes and only one eigenvalue of $\tau$
is different from zero. Generally the number of eigenvalues of  $\tau$
which differ from zero is less or equal than $M_<(\epsilon)$, where  $M_<(\epsilon)$ is
the smaller of the two $M_a(\epsilon)$. This stems from the fact that
$AA^\dagger$ and $A^\dagger A$ have the same nonvanishing eigenvalues
with the same multiplicities.
For $M_L(\epsilon)<M_R(\epsilon)$ one better calculates the eigenvalues of
$A^\dagger A$ which is a hermitian $M_L(\epsilon)\times M_L(\epsilon)$ matrix.\\

In order to compare our general result with the Levitov-Lesovik
formula one can use
\begin{eqnarray}
e^{-i\lambda}(1+d(\lambda)X_{\mu,+})(1+d(\lambda)X_{\mu,-})~~~~~~~~~
~~~~~~~~~~~~~~~ \nonumber \\
~~~~=e^{-i\lambda}\left[1+ (1+(f_L-f_R)\tau_\mu)d(\lambda)
+\tau_\mu \bar f_Rf_L  d(\lambda)^2     \right] \nonumber \\
=1+\tau_\mu (d(\lambda) f_L\bar f_R +d(\lambda)^*f_R\bar
f_L)~.~~~~~~~~~~~~~~~~~~~~~
\end{eqnarray}
With ${\rm det}(1+d(\lambda)\tilde
X_\alpha(\epsilon))=\prod_\alpha (1+d(\lambda)X_\alpha(\epsilon ))$
this yields
\begin{eqnarray}
\label{newresult}
\ln g_R\approx \frac{t}{2\pi} 
\int_{\epsilon_{\rm min}}^{\epsilon_{\rm max}} \sum_{\mu=1}^{M_<(\epsilon)}
\ln \left[ 1+\tau_\mu (d~
f_L\bar f_R +d^*f_R\bar
f_L) \right ]d\epsilon.~
\end{eqnarray}
 This completes the derivation of the generalization of the
 Levitov-Lesovik formula for two general quasi-one-dimensional
 leads \cite{comment2}.
 The integration has to be split up into $N_I$ energy intervals
from $\epsilon^{(i)}$ to $\epsilon^{(i+1)}$ where $i=1,...,N_I$ and 
$\epsilon^{(1)}=\epsilon_{\rm min}$ and $\epsilon^{(N_I+1)}=\epsilon_{\rm
  max}$. In the intervals the number of open channels
determined by the  $M_a(\epsilon)$ is constant. We denote the constant
value of $M_<(\epsilon)$ in the $m$-th interval by $M_<^{(m)}$.
  The splitting is discussed in the following section for simple model
 system.

With the eigenvalues $X_\alpha(\epsilon)$ of the $M(\epsilon)\times
M(\epsilon)$
 matrix $\tilde X(\epsilon)$ and the
transition from the integral to a finite Riemann sum as in
Eq. (\ref{Riemannsum}) one can obtain an approximation for the
eigenvalues of the operator $\tilde X(t)$ defined in
Eq. (\ref{Xtilde}) 
which generalizes the introduction of the $a_\pm$ in section III. The 
approximation for $g_R$ reads
\begin{eqnarray}
g_R(t,\lambda)\approx \prod_{I_m}e^{-i\lambda 2 M^{(m)}_<
  N_m(t)}
\prod_{j=1}^{N_m(t)}\prod_{\alpha=1}^{2M^{(m)}_<}\left [1+d(\lambda)
 X_\alpha(\epsilon^{(m)}_j )\right] 
\end{eqnarray}
with $N_m(t)=(\epsilon^{(m+1)}- \epsilon^{(m)})t/2\pi$
and the energy variables are given by $\epsilon_j^{(m)}=\epsilon^{(m)}+2\pi(j-1/2)/t$.
Therefore the approximation for the eigenvalues $\tilde X_m(t)$ which
differ from $1$ and $0$ are given by the $ X_\alpha(\epsilon_j^{(m)})$.
The time dependence enters via the $\epsilon_j^{(m)}$.

\section{Almost perfect stripe}

In order to elucidate our general result Eq. (\ref{newresult}) we
consider leads which are both stripes of width $N_{\perp}$.  Analytical
results for the eigenvalues $\tau_\mu$ are presented for the case
where the dot region is identical to the leads except for a single
site impurity
\begin{eqnarray}
\label{almost}
h=-\sum_{m=-\infty}^\infty
\sum_{n=1}^{N_\perp-1}&[&t_{\|}|m,n\rangle\langle m+1,n|+
t_\perp  |m,n\rangle\langle m,n+1| \nonumber \\
&+&H.c   ]
+V_0|0,n_0\rangle\langle0,n_0| 
\equiv \tilde h_0+\tilde v_0 ~.   
\end{eqnarray}
The tildes are introduced to indicate that the separation of the
Hamiltonian  in the
unperturbed part and the perturbation is different from the one 
used in Eq.(\ref{Hamiltonian}) and Eqs. (\ref{hamiltonian}) and
(\ref{connection}).
 
For $V_0=0$ this is an ideal infinite stripe with eigenvalues
\begin{equation}
\epsilon_{k,l}=-2t_{\|}\cos{k}-2t_\perp
\cos{\frac{l\pi}{N_\perp+1}}\equiv
\epsilon_k^{\|}+\epsilon_l^\perp~,
\end{equation}
with $k\in [-\pi,\pi]$ and $l=1,...,N_\perp$.
For a given energy $\epsilon$ there are $2N_\perp(\epsilon)\le 2N_\perp$
scattering channels open for which $|\epsilon-\epsilon_l|\le 2|t_\||$ holds.
 The standing wave lead states
$|\epsilon,a,l\rangle$ are labeled by $a=L,R$ and the transverse quantum
numbers $l$ of the open channels.
 The scattering matrix is obtained
via the $t$-operator \cite{Taylor}
\begin{equation}
t(z)=v+vg(z)v~,
\end{equation}
where $g(z)=(z-h)^{-1}$ is the exact resolvent and $v$ is the
generalization of the operator connecting the leads with the dot in
Eq. (\ref{connection}). For the almost perfect stripe it is given by
\begin{eqnarray}
v&=&-t_{\|}\sum_{n=1}^{N_\perp}(|-1,n\rangle\langle
0,n|+|0,n\rangle\langle1,n|+H.c) \nonumber \\
&=&-t_{\|}\sum_{l=1}^{N_\perp}(|l_{(-1)}\rangle\langle
l_{(0)}|+|l_{(0)}\rangle\langle l_{(1)}|+H.c) ~,
\end{eqnarray}
where the $|l_{(m)}\rangle$ are the standing wave eigenstates in the
perpendicular direction formed from the states $|m,n\rangle$. Then the
$t$-matrix elements take the simple form
\begin{equation}
\label{tmatrix}
\langle
\epsilon,a,l|t(z)|\epsilon,a',l'\rangle=t_{\|}^2\langle
\epsilon,a,l|l_{(a)}\rangle \langle l_{(0)}|g(z)|
l'_{(0)}\rangle
\langle l'_{(a')}|\epsilon,a',l'\rangle,
\end{equation}
where $|l_{(a)}\rangle $ is the standing wave state at $n=-1$ for
$a=L$ and  $n=1$ for
$a=R$.
The $t$-matrix elements enter the scattering matrix for
$z=\epsilon+i0$.

 The exact
resolvent matrix elements can easily be calculated
for the Hamiltonian in Eq. (\ref{almost}) as the site impurity
provides a separable perturbation. With $\tilde g_0(z)=(z-\tilde
h_0)^{-1}$
one obtains
\begin{eqnarray}
\label{resolvent}
\langle l_{(0)}|g|l'_{(0)}\rangle &=& \langle
l_{(0)}|\tilde g_0|l'_{(0)}\rangle \\
&+&  \frac{
\langle l_{(0)}|\tilde g_0|0,n_0\rangle
V_0 \langle 0,n_0|\tilde g_0|l'_{(0)}\rangle  }
{1-V_0 \langle
  0,n_0|\tilde g_0|0,n_0\rangle }.
\nonumber
\end{eqnarray}
For the open channels 
\begin{equation}
\langle l_{(0)}|\tilde g_0(\epsilon+i0)|l'_{(0)}\rangle
=\delta_{ll'}\frac{-i}{\sqrt{B_{\|}^2-(\epsilon-\epsilon_l)^2}}
\end{equation}
holds, with $B_{\|}=2t_{\|}$. The $\tilde g_0$-matrix element in the
denominator in Eq. (\ref{resolvent})
involves contributions from the open and the closed channels
\begin{eqnarray}
\langle 0,n_0|\tilde g_0(\epsilon+i0)|0,n_0\rangle&=& \sum_{l (open)} 
\frac{-i |\langle
  l_{(0)}|0,n_0\rangle|^2}{\sqrt{B_{\|}^2-(\epsilon-\epsilon_l)^2}} \\
&+& \sum_{l (closed)} 
\frac{ |\langle
  l_{(0)}|0,n_0\rangle|^2}{\sqrt{(\epsilon-\epsilon_l)^2-B_{\|}^2
  }}\nonumber\\
&\equiv& -i\pi\tilde \rho_{00}+\tilde g_{00}^R~. \nonumber
\end{eqnarray}
While the contribution of the open channels is purely imaginary the
one of the closed channels is real. To complete the calculation of the
$t$-matrix elements in Eq. (\ref{tmatrix}) the overlaps $\langle
\epsilon,a,l|l_{(a)}\rangle $ are needed. They are related to the density of
states at the boundary of a semi-infinite chain
\begin{equation}
t_{\|}\langle
\epsilon,a,l|l_{(a)}\rangle=\left(\frac{1}{2\pi}\sqrt{B_{\|}^2-(\epsilon-\epsilon_l)^2}
                   \right)^{1/2}
\end{equation}
This leads to the scattering matrix
\begin{equation}
s_{al,a'l'}(\epsilon)=\delta_{ll'}(\delta_{aa'}-1)+(l|n_0)u(n_0|l')
\end{equation}
with
\begin{equation}
(l|n_0)=\frac{ \langle
  l_{(0)}|0,n_0\rangle}{\sqrt{B_{\|}^2-(\epsilon-\epsilon_l)^2}
}~,~~~~u=\frac{iV_0}{1-V_0\tilde g_{00}^R+i\pi V_0\tilde \rho_{00}}.
\end{equation}
With the $N_\perp(\epsilon) \times N_\perp(\epsilon)$ projected
scattering matrix $s_{RL}=-1+|n_0)u(n_0|$ the operator $\tau $ defined
in Eq. (\ref{deftau}) is given by
\begin{equation}
\tau=1-(u+u^*-(n_0|n_0)|u|^2)|n_0)(n_0| ~.
\end{equation}
Because of the separable form of $\tau-1$ the only eigenvalue of
$\tau$ different from $1$ is given by
\begin{equation}
\tau_1=1-(u+u^*-(n_0|n_0)|u|^2)(n_0|n_0)~.
\end{equation}
Using $(n_0|n_0)=\sum_{l (open)}(n_0|l)(l|n_0)=\tilde \rho_{00}/2 $
one finally obtains
\begin{eqnarray}
\tau_1&=&\frac{(1-V_0 \tilde g_{00}^R)^2}{(1-V_0 \tilde g_{00}^R)^2+(\pi V_0
 \tilde \rho_{00})^2 }~, \\
\tau_2&=&\tau_3=...=\tau_{N_\perp(\epsilon)}=1~. \nonumber
\end{eqnarray}
The ``perfect transmission'' eigenvalues $\tau_i=1$ yield
contributions to $\ln{g_R}$ of the form discussed following
Eq. (\ref{pt}). As the energy integrations in Eq. (\ref{newresult})
for fixed $M_<(\epsilon)$
are over restricted energy ranges the logarithmic 
corrections in Eq. (\ref{pt}) are important. 

The simplest case is a $N_\perp=2$ ladder system. For the special case
$t_\|=t_\perp=1$ there are two bands corresponding to
$\epsilon^\perp_l=\pm 1$ of width 4. Therefore one has to distinguish
the energy intervals $[-3,-1],[-1,1]$ and $[1,3]$ with one, two and one
open channel. When both channels are open  $\tau_2=1$ and  $\tilde g^R_{00}$ vanishes.
This implies $\tau_1\to 0$ for $V_0\to \infty$, while $\tau_1$ stays finite in
this limit when only one channel is open.

\section{Summary}

In this paper we have generalized the exact numerical method to
calculate the FCS for large but finite systems \cite{KS} to finite
temperatures using the eigenvalues of the operator $X(t)$ in the
Hilbert space of a single particle introduced by Abanov and Ivanov\cite{AI}. 
In the long time limit  the results for the
probability distribution for the number of transmitted particles
agree well with the result using the
Levitov-Lesovik approximation \cite{LL1,LL2}
except for the case of (almost) perfect transmission.

Using a similar identity for the finite temperature leading order in
time result for the logarithm of the characteristic function a new
explicit result for $\ln{g_R}$ was presented in Eq. (\ref{newresult}) for two general
quasi-one-dimensional leads which involves the eigenvalues of a matrix
formed from the transmission amplitudes. For a simple
model these eigenvalues were calculated analytically.

\ack
The author would like to thank W. Zwerger for the hospitality during
his sabbatical stay at the TU M\"unchen.

\begin{appendix}
\section*{Appendix}
\setcounter{section}{0}

\section*{Recursive step in the  calculation of  $w_R(t,Q)$}

The numerical finite temperature results presented in section 3 were
obtained by first calculating the eigenvalues $\tilde X_m(t)$ and then
performing the product in Eq. (\ref{AbIv}). This is done iteratively
as follows.

 Let $F_N(x)$ be a polynomial given in the form of a
product
\begin{equation}
F_N(x)=\prod_{i=1}^N(a_i+b_i x)=\sum_{m=0}^Nc_m^{(N)}x^m
\end{equation}
The coefficients $c_m^{(N)} $ are obtained iteratively
by calculating the polynomials $ F_M(x)$ with
coefficients $c_m^{(M)}$ starting with $M=1$ and using
\begin{eqnarray}
F_{M+1}(x)&=&(a_{M+1}+b_{M+1}x)F_M(x) \\
&=& \sum_{m=0}^M(a_{M+1}c_m^{(M)}x^m+b_{M+1}c_m^{(M)}x^{m+1})\nonumber
\end{eqnarray}
This leads to the recurrence relations
\begin{eqnarray}
c_0^{(M+1)}&=&a_{M+1}c_0^{(M)}  \\
c_m^{(M+1)}&=&a_{M+1}c_m^{(M)}+b_{M+1}c_{m-1}^{(M)}~,~~   1\le m \le M
\nonumber \\
c_{M+1}^{(M+1)}&=&b_{M+1}c_M^{(M)}. \nonumber
\end{eqnarray}

\end{appendix}


\begin{thebibliography}{99}
\bibitem{comment} The matrix $A$ is usually denoted as $t$ \cite{BB}.
We chose not to called it $t$ as we use this letter for the
$t$-operator.
\end{thebibliography}
\end{document}